\documentstyle[preprint,epsf,aps]{revtex}
\begin{document}
\draft

\title{Direct radiative capture of p-wave neutrons} 

\author{A. Mengoni,$^{1,2}$ T. Otsuka,$^{2,3}$ 
and M. Ishihara$^{2,3}$}

\address{$^{1}$ENEA, Applied Physics Section, 
V.le G. B. Ercolani 8, I-40138 Bologna, Italy \\
$^{2}$RIKEN, 
Radiation Laboratory, 2-1 Hirosawa, Wako, Saitama 351-01, Japan \\
$^{3}$The University of Tokyo, Department of Physics, 
Hongo, Bukyo-ku, Tokyo 113, Japan }

\date{\today}
\maketitle
\begin{abstract}

The neutron direct radiative capture (DRC) process is 
investigated, highlighting 
the role of incident $p$-wave neutrons.
A set of calculations is shown for the
$^{12}$C$(n,\gamma)$ process at incoming neutron
energies up to 500 keV, a crucial region for astrophysics. 
The cross section for neutron
capture leading to loosely bound $s$, $p$ and $d$ orbits
of $^{13}$C is well reproduced by the DRC model demonstrating
the feasibility of using this reaction channel to study
the properties of nuclear wave functions on and outside the
nuclear surface.
A sensitivity analysis of the results on the neutron-nucleus 
interaction is performed for incident $s$- as well as $p$-waves.
It turned out that the DRC cross section for  
$p$-wave neutrons is insensitive to this interaction, 
contrary to the case of incident $s$-wave neutrons.

\end{abstract}


\narrowtext

The direct radiative capture (DRC) process of neutrons
in the keV energy region has some peculiarity
recently revived by theoretical analysis \cite{Otx94} 
as well as by new experimental results \cite{Nax91,Ohx94,Igx95}. 
Because of the non-resonant nature of the DRC process, 
the complications related to the
calculation of the compound nucleus wave function
in the entrance channel are removed.
This is a general feature of all the
direct capture processes, including those induced
by charged particle reactions. However, because
of the lack of the Coulomb interaction, the
$(n,\gamma)$ reaction has salient features
which makes it a unique probe for
investigating nuclear structure information.
In fact, the neutron capture process can be explored 
in the very low neutron energy region where the reaction 
mechanism may be fully decoupled from the resonance process.
In this way, precise information can be obtained for
the structure of the capturing orbit
and the relative contribution of the 
various $l$-wave components to the 
cross section can be examined separately.

Because the DRC process is 
essentially taking place on the nuclear
surface and in the external region, it has been recently 
proposed \cite{Otx94} to use this reaction channel 
to study the properties of nuclear wave functions, 
in connection with the discovery of the neutron halo 
of light drip-line nuclei \cite{Tax85Ta91}.
The same kind of information can also
be derived from the inverse reaction channel
(Coulomb dissociation) where a strong enhancement
of the low-lying dipole mode has been
observed \cite{Nax94} and treated as 
an inverse DRC process \cite{Otx94}.

While the DRC of protons and alpha particles
have been widely investigated in the
energy range from a few hundreds of keV
up to several MeV \cite{Kix94}, the DRC process of
neutrons has been mainly examined at 
thermal ($E_n = 0.0253$ eV) 
energies where $s$-wave neutrons 
are captured into bound $p$ orbits. 
The DRC formalism for 
thermal neutrons has been revised by 
Raman {\it et al.}\cite{Rax85}. 
They have shown in detail how the neutron-nucleus 
potential strongly affects the capture mechanism 
of $s$-wave neutrons in light nuclei, whereas no 
reference to high energy extension nor to higher 
partial-wave (including $p$-waves) contributions was given. 
On the other hand, we may expect that, as the incoming 
neutron energy increases
the capture of $p$-wave neutrons into bound
$s$ and $d$ orbits comes into play
and, under proper conditions, it can be regarded 
as the dominant capture process. 

The extension of the capture models required to 
include $p$-waves and higher partial waves into the 
neutron DRC process is the main task of this note. 
In addition, a sensitivity 
analysis of the DRC process to the neutron-nucleus
potential for energies in the keV region will
be performed.
We stress that applications of DRC models have been 
rarely extended to energies higher than thermal. In nuclear 
astrophysics there have been several applications at neutron 
energies of interest in the {\it r}-process nucleosynthesis 
\cite{Max83} and for inhomogeneous big-bang 
theories \cite{Wix90}. For such kind of applications 
it is necessary to assess quantitatively the
DRC prescriptions in a range of energies from a few
up to several hundreds of keV.

Here we will briefly revise the DRC model for $s$- and 
$p$-wave neutrons and apply it to the calculation of the
$^{12}$C$(n,\gamma)$ cross sections for transitions
leading to all the four bound states of $^{13}$C .
In particular we will consider realistic
wave functions for the initial scattering state and
we will focus the attention on the influence of the 
initial $l$-wave character on the capture cross section.


In the early works on neutron capture 
reactions \cite{Th51,MI60,LL60,Ly68} it was recognized 
that a capture mechanism, in which the incoming neutron 
is scattered directly into a final bound state without 
forming a nuclear compound state, might take place for nuclei 
where the final state is dominated by a strong 
single-particle configuration.
There have been several formulations 
of the DRC mechanism differing considerably among each 
other in the way the incoming channel and the final 
state are described \cite{LL60,LM74,Rax85,Otx94}.
In general, because the direct capture process is alternative 
to the compound nucleus (CN) formation mechanism,
we can separate the collision matrix into two components
\begin{equation}                                                        
U_{i \rightarrow f} = U_{i \rightarrow f}(CN) + 
U_{i \rightarrow f}(DRC)
\end{equation}
where all the quantum numbers
necessary to define the initial and final states
have been lumped into the notation $i$ and $f$, respectively.
The reaction cross section is given by
\begin{equation}                                                        
\label{eq3}                                                          
\sigma_{i \rightarrow f} = \frac{\pi}{ {k}^2 } 
\vert U_{i \rightarrow f} \vert ^2
\end{equation}                                                          
where $k$ is the wave number of the relative motion in the
entrance channel. 
Here we will deal only with the DRC part of the collision
matrix. The capture cross section 
for emission of electric dipole radiation (E1)    
in the transition $i \rightarrow f$ is given by 
\begin{equation}                                                        
\sigma_{n,\gamma} = \frac{16 \pi}{9 \hbar} k_{\gamma}^{3} \bar{e}^2 
\vert {\it Q^{(1)}_{i \rightarrow f}} \vert^{2}         
\end{equation}
where $k_{\gamma} = \epsilon_{\gamma} / \hbar c$ 
is the emitted $\gamma$-ray wave number 
corresponding to the $\gamma$-ray energy $\epsilon_{\gamma}$ 
and $\bar{e} = -eZ/A$ is the E1 effective
charge for neutrons. The cross section is, therefore,
essentially determined by the matrix elements
\begin{equation}                                                        
{\it Q^{(1)}_{i \rightarrow f}} = < \Psi _{f} 
\vert \hat{T}^{E1} \vert \Psi _{i} >
\end{equation}
where
%
$ \hat{T}^{E1} = r Y^{(1)}(\theta,\phi)$
is th electric dipole operator.
Here, the initial state wave-function $\Psi _{i}$ is given by
a unit-flux incoming wave in the entrance channel, scattered at the origin
by the neutron-nucleus potential. The final state wave-function 
$\Psi _{f}$ is given by the residual nucleus (bound) final state.
The radial coordinate $r$ denotes the distance of the 
incoming neutron with respect to the target nucleus.

The entrance channel wave function can be decomposed
into spherical ($l$-wave) components
\begin{equation}                                                        
\Psi _{l m}(\mbox{\bf r}) \equiv  
w_{l}(r)
\frac{Y_{l,m}(\theta,\phi)}{r v^{1/2}}
\end{equation}
where $w_{l}(r)$ depends also on the 
wave number $k$ and is written, as usual, as
\begin{equation}                                                        
w_{l}(r) = \frac{i \sqrt{\pi}}{k}
\sqrt{2l+1} i^{l}
[I_l - U_{l} O_l].
\end{equation}
Here, the common notation for the asymptotic forms 
of the incoming and outgoing waves, respectively
$I_l$ and $O_l$, has been adopted
\begin{equation}                                                        
I_{l} \sim \mbox{exp}( -ikr + \frac{1}{2}il\pi ) 
\quad \mbox{and} \quad
O_{l} \sim \mbox{exp}( +ikr - \frac{1}{2}il\pi ). 
\end{equation}                                                          
$U_{l}$ indicates the collision matrix for the
scattering process in the entrance channel,
$v$ is the incoming neutron velocity and $k$ the 
corresponding wave number. 

The matrix elements can be decomposed into the 
product of three factors
${\it Q^{(1)}_{i \rightarrow f}} = {\cal I}_{if} \cdot A_{if} \cdot \sqrt{S} $.

\vspace{0.2 cm}
\noindent{\it Radial part: }
Indicating with $u_{l_f}(r)$ the radial part of the final 
state wave function, the radial overlap integral is given by
\begin{equation}                                                        
\label{eqI}
{\cal I}_{l_{i}l_{f}} \equiv \int_{0}^{\infty} 
u_{l_f}(r) r w_{l_i}(r) dr. 
\end{equation}
If the final state wave function is dominated 
by a single-particle configuration with a long tail outside
the nuclear radius (halo), there will be a strong effect 
on ${\cal I}_{l_{i}l_{f}}$ \cite{Otx94}. This will be shown below in 
the case of incoming $s$- and $p$-wave neutrons.
Analytical expressions for ${\cal I}_{l_{i}l_{f}}$ can be derived
for specific assumptions on the initial and final state
wave functions. If a hard-sphere model for
the scattering wave function and a crude square-well
model for the bound $p$ orbital are assumed for
the initial and final state respectively one recovers the very
well known\cite{LL60} expression for the hard-sphere 
capture cross section
\begin{equation}                                                        
\label{eq14}                                                          
\sigma_{n,\gamma}^{HS} = \frac{32\pi}{3} \enspace
k_{\gamma}^{3} \enspace
\frac{\bar{e}^2}{\hbar v} \frac{R^5}{y^4}
(\frac{3+y}{1+y})^2
\end{equation}                                                          
where $y \equiv \chi R$. Here, $\chi$ is the
reciprocal attenuation length of the wave function
tail given by $\chi = \sqrt{2 \mu S_n}/ \hbar$,
where $\mu$ is the reduced mass of the system and
$S_n$ the neutron separation energy from the
residual state. 
More elaborate expressions corresponding to
different assumptions on the initial and
final state wave functions, can be derived for this
overlap integral and they can be found in the
literature \cite{LL60,Rax85,Mex94}. 

\vspace{0.2 cm}
\noindent{\it Angular part: }
For a general $l_{i} \rightarrow l_{f}$ transition, the
angular factor which takes into account the magnetic
degeneracy of the final state and is averaged over the 
initial magnetic substates is given by
\begin{equation}                                                        
\label{eq20}                                                          
A_{l_i l_f}^{2} \equiv \frac{1}{2l_{i} + 1} 
\vert \langle l_{f} \| \hat{Y}_{1} \| l_{i} \rangle \vert ^2 =
\frac{3}{4 \pi} (l_i 0 10 \vert l_f 0 )^2
\end{equation}                                                          
where $(.... \vert ..)$ is a Clebsh-Gordan coefficient.
The orbital angular momentum, $\bf{l}$, can couple with the intrinsic 
spin, $\bf{s}$, to give the angular momentum
${\bf j} = {\bf l} + \bf{s}$. 
The corresponding 
${\bf j}_i$ and ${\bf j}_f$ 
can be combined with the target spin 
${\bf \mbox{I}}$ 
to obtain the initial and final state total angular momenta 
${\bf J}_i$ and ${\bf J}_f$.
The angular-spin coefficient for the full coupling is given by
\widetext
\begin{eqnarray}
A_{if}^{2} & \equiv & 
\frac{ 
\vert \langle J_{f}, I, j_{f} 
\| \hat{Y}_{1} \| 
J_{i}, I, j_{i} \rangle \vert ^2}
{2J_{i} + 1} =  
\frac{3}{4 \pi} (2l_{i}+1)(2j_{i}+1)(2j_{f}+1)(2J_{f}+1)
\times \nonumber \\ 
           & \times & (l_i 0 10 \vert l_f 0 )^{2} \enspace
\{ \begin{array}{ccc}
j_{i} & I & J_{i} \\
J_{f} & 1 & j_{f} \end{array} \} ^2
\{ \begin{array}{ccc}
l_{i} & 1/2 & j_{i} \\
j_{f} &  1  & l_{f} \end{array} \} ^2
\label{eq21}                                                          
\end{eqnarray}                                                          
\narrowtext
where $\lbrace ...\rbrace$ are Wigner-6j coefficients.

\vspace{0.2 cm}
\noindent{\it Final state strength: }
The single particle strength, $S$, 
of the final-state orbit is usually derived from $(d,p)$ 
stripping reactions (spectroscopic factor).
We note here that if the radial part of the matrix 
elements, Eq.\ (\ref{eqI}), is calculated with reliable wave 
functions (see below)
the spectroscopic factor can be derived from a DRC 
analysis of the experimental cross section. This
technique, proposed and applied in proton capture
reactions \cite{Kix94}, could be applied in the neutron
capture channel with the advantages already
noted above.

For thermal neutrons, the main contribution to the
capture process is due to incoming $s$-wave neutrons 
captured into $p$-wave orbits and emitting E1 radiation.
At higher neutron energy, however, the incoming $p$-wave
neutrons come into play as they can be captured into
bound $s$ and $d$ orbits. In $^{13}$C both these
even parity orbits are available. Moreover their
respective levels have the characteristics of being loosely 
bound and with large spectroscopic strengths: they are 
good example of halo structure in excited states
of stable nuclei.
The nuclear structure information on $^{13}$C are 
summarized in Table \ref{tab1}.

Before showing the result of the full $^{12}$C$(n,\gamma)$ 
cross section calculations we will show here how
the $s$- and $p$-wave neutron capture is affected
by the neutron-nucleus potential.

For the purpose of the present study it is
sufficient to consider single particle wave functions
for the bound-state orbits of $^{13}$C. They
have been calculated using a Wood-Saxon potential with 
a radius parameter $r_0$ = 1.236 fm, 
a diffuseness $d$ = 0.62 fm and 
a spin-orbit potential strength $V_s$ = 7 MeV. 
The potential well depths were adjusted so as to
reproduce the correct binding of the four bound states
in terms of the corresponding single particle orbits. 
In the case of the $2s_{1/2}$ state this gives 
rise to $V_0$ = 59.23 MeV.

In the present investigation we have treated the
incoming neutron channel with the following 
approximations:

\begin{itemize}
\item[(a)] plane wave (PW) approximation
$$
\begin{array}{llll}
U_{l} & = & 1  &  \qquad \qquad 
\mbox{for all {\it l}} \\
\end{array}
$$

\item[(b)] hard-sphere (HS): scattering by an 
infinitely deep potential well of radius $R$ 

$$
\begin{array}{llll}
U_{l} & = & e^{ -2i k R } & \qquad \qquad 
\mbox{for {\it l} = 0} \\
      & = & e^{ -2i k R } \times \frac{1+ik R}{1-ik R} & 
            \qquad \qquad \mbox{for {\it l} = 1} \\ 
\end{array}
$$

\item[(c)] a general case in which the collision 
matrix is calculated  numerically for a given potential of 
Wood-Saxon form.
\end{itemize}

\begin{figure}
\begin{center}
\leavevmode
\hbox{%
\epsfxsize=12.6cm \epsffile{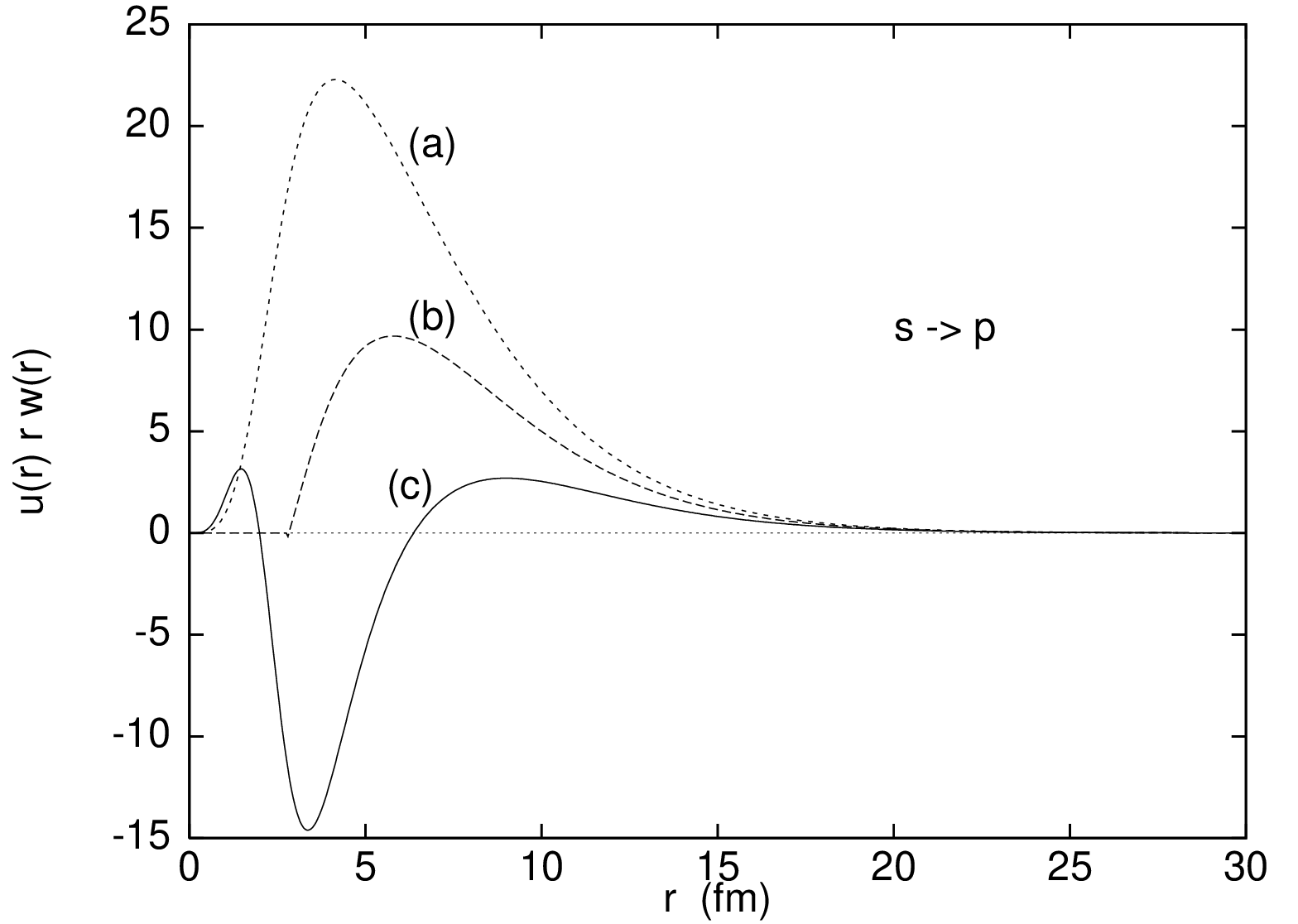}}
\end{center}
\caption{Real part of the integrand of Eq.\ (\protect\ref{eqI}) 
for incoming $s$-wave neutrons of 200 keV and
for the $1p_{3/2}$ orbit bound by 4.96 MeV. 
The scattering wave function $w(r)$ has been calculated according 
to (a) PW, (b) HS and (c) Wood-Saxon potential prescriptions. 
See the text for parameters and a detailed explanation.}
\label{fig1}
\end{figure}

\begin{figure}
\begin{center}
\leavevmode
\hbox{%
\epsfxsize=12.6cm \epsffile{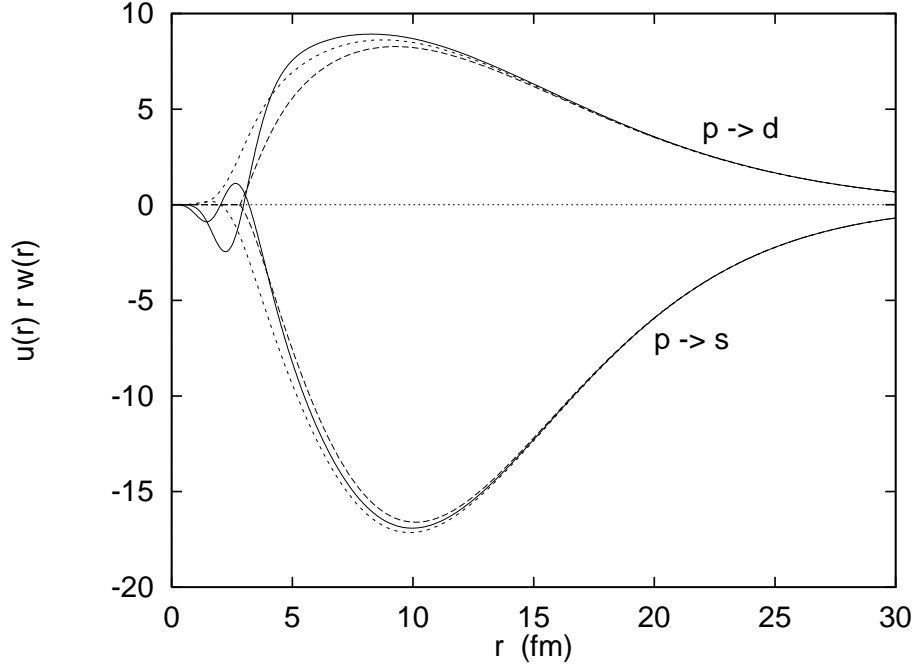}}
\end{center}
\caption{Imaginary part of the integrand of Eq.\ (\protect\ref{eqI}) 
for incoming $p$-wave neutrons of 200 keV and
for the $2s_{1/2}$ and $1d_{5/2}$ orbits
bound by 1.86 and 1.09 MeV respectively. 
The scattering wave functions $w(r)$
have been calculated according to a PW (dashed line), HS (dotted line)
and Wood-Saxon potential prescriptions (solid line). 
See the text for parameters and a detailed explanation.}
\label{fig2}
\end{figure}

In order to see the effect of these three
different treatments on the radial part
of the matrix elements we show 
in Figs.~\ref{fig1} and ~\ref{fig2} 
the calculation of the 
integrand of ${\cal I}_{if}$ in Eq.\ (\ref{eqI})
at $E_n$ = 200 keV.
Because of its dominant contribution to the
integral, only the real (imaginary) part is shown in
Fig.~\ref{fig1} (~\ref{fig2}) for $s$- ($p$-) wave
capture. The cross section is proportional
to the squared area under the curves shown in the 
figures.

From Fig.~\ref{fig1} it is evident that 
for an initial $s$-wave state the result 
is strongly dependent on the neutron-nucleus 
potential adopted, namely on the different collision 
matrix of the scattering channel used in the calculation.
Moreover, the behavior of the wave function 
inside the nuclear radius may result in a significant 
cancelation of the radial matrix element.
This result is in full agreement with the conclusions
of a detailed study of the thermal capture 
in light nuclei by Lynn {\it et al.} \cite{Lyx87}. 
We believe that this sensitivity is the main source 
for discrepancy observed between the calculated 
and experimental thermal cross sections in DRC model
analysis. 
The influence of compound nucleus components 
in the collision process is, naturally, another source 
of uncertainty. 

On the contrary, for incoming $p$-wave neutrons,
the radial matrix elements are essentially insensitive 
to the different collision matrices considered.
This is evident, from Fig.~\ref{fig2}, for both the
$p \rightarrow s$ and the $p \rightarrow d$ transitions. 
In other words, in a situation where the capture
is dominated by a DRC process of $p$-wave neutrons, the
cross section is not sensitive to the neutron-nucleus
potential. Hence, the capture process is essentially
determined by the structure of the {\it final} state
wave function. In particular, the component of the
wave function outside the nuclear radius plays the
principal role in the determination of the capture
strength. This is the main result of the present 
investigation.

\begin{figure}
\begin{center}
\leavevmode
\hbox{%
\epsfxsize=12.6cm \epsffile{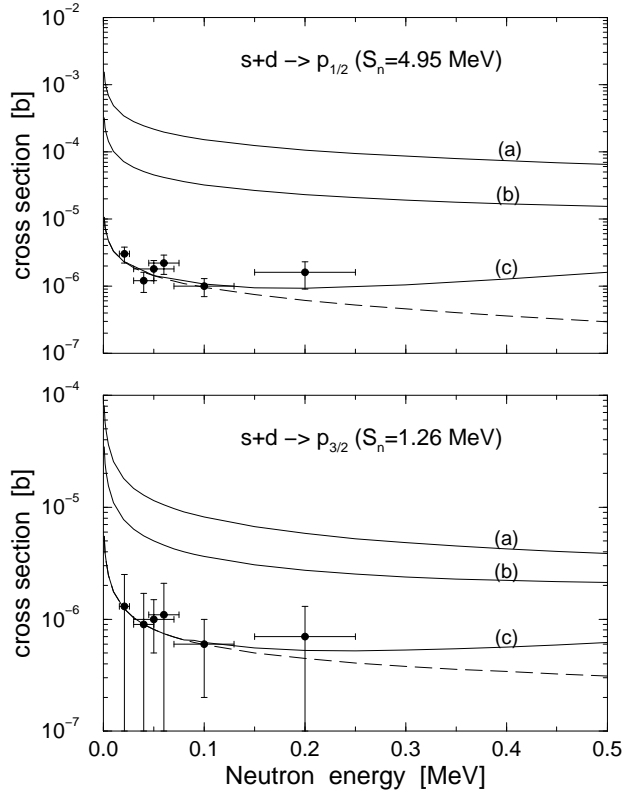}}
\end{center}
\caption{Neutron capture cross section of $^{12}$C for 
transitions leading to the ground state of $^{13}$C 
({\it upper part}) and to the
level at $E_x$ = 3.684 MeV ({\it lower part}). 
The experimental values are from the reference \protect\cite{Nax91}.
The curves labeled with (a), (b) and (c) correspond to the three
different assumptions concerning the collision matrix for the
scattering channel. 
See Figs.\ (\protect\ref{fig1}) and
text for explanation. The dashed-line represents exclusively
$s$-wave contribution to the capture.}
\label{fig3}
\end{figure}

\begin{figure}
\begin{center}
\leavevmode
\hbox{%
\epsfxsize=12.6cm \epsffile{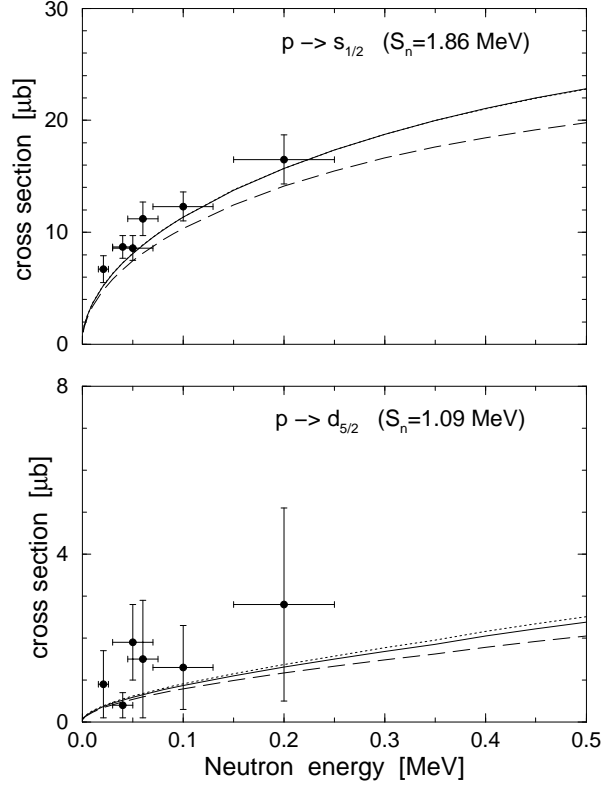}}
\end{center}
\caption{Neutron capture cross section of $^{12}$C for 
transitions leading to the $s_{1/2}$ orbit ({\it upper part}) 
and to the $d_{5/2}$ orbit of $^{13}$C ({\it lower part}). 
The experimental values are from Ref. \protect\cite{Nax91}.
The dashed-, dotted- and solid-line correspond to the three
different assumptions concerning the collision matrix for the
scattering channel. 
See Fig.\ (\protect\ref{fig2}) and
text for explanations. In this case, the capture is due
to incoming $p$-wave neutrons only.}
\label{fig4}
\end{figure}

The results of the full calculation of the capture
cross section for transitions leading to the four
bound states in $^{13}$C are shown in 
Figs.~\ref{fig3} and ~\ref{fig4}.

In Fig.~\ref{fig3}, the capture cross sections for 
transitions leading to the ground state (upper part) 
and to the level at $E_x$ = 1.26 MeV (lower part) 
are shown. 
The curves labeled by (a), (b) and (c) refer
to three different scattering wave functions
calculated according to the three treatments  
(PW, HS and diffused Wood-Saxon potential)
described above. For these two transitions, 
incoming $s$- and $d$-wave neutrons are involved 
in composing the initial state wave function.
The large discrepancy between the experimental values 
and the calculation is removed only if a realistic
neutron-nucleus potential is employed in the
calculation. In our case, such an agreement was
obtained when the same potential used for the 
bound-state calculation was employed (curve (c) in 
the figure). This is an important point because 
the obtained wave functions might have had unphysical 
overlaps and should have been orthogonalized if different 
potentials would have been used.

We have extended the calculations down to thermal
energy ($E_n$ = 0.0253 eV) using the same  
Wood-Saxon potential and the results are shown
in Table \ref{tab1}. Considering that no adjustment
of any parameter was performed, the results of our
DRC calculation can be considered satisfactory.

The relative contribution of incoming $d$-waves
can be seen in Fig.~\ref{fig3}. There, the dotted 
lines show the contribution of the $s$-wave capture 
component only, whereas the full line (c) show the 
contribution of both $s$- and $d$-wave components. 
Thought not decisively, the experimental 
results seem to follow the increasing trend due 
to the onset of the higher partial wave component.

Finally, the cross sections for transitions leading to
the $s_{1/2}$ and $d_{5/2}$ orbits are shown 
in Fig.~\ref{fig4}.
These are the transitions due to the capture from
$p$-wave neutrons. 
A very good agreement is found for both the
reaction channels, implying reliability of the
DRC mechanism in the energy region under consideration.
In this figure the conclusion drawn above from the radial
matrix elements calculations can be verified explicitly.
The results of the calculations obtained using the PW,
HS or the full Wood-Saxon potential are barely
distinguishable. No major influence of the
neutron-nucleus potential used is revealed by the
calculations, very well supported by the experimental 
results.

In summary, our calculations have shown that, while
the DRC process of $s$-wave neutrons is strongly
biased by the incoming-neutron interaction with
the target, the DRC of $p$-wave neutrons is essentially
insensitive to the details of this interaction.
This fact can be used to derive important nuclear
structure information on the residual nucleus like
``exotic'' components of the neutron wave function 
outside the nuclear radius (i.e. neutron halo) or single 
particle strength (spectroscopic factor) of the final 
capturing state.

\vspace{1.0pc}
We acknowledge many fruitful discussions with 
Y. Nagai and C. Coceva. 
This work has been supported in part by Grant-in-Aid for 
Scientific Research on Priority Areas (No. 05243102).


\clearpage

\clearpage

\clearpage

\clearpage

\clearpage
\begin{table}
\begin{center}
\caption{Nuclear structure information of $^{13}$C and thermal
neutron capture cross section of $^{12}$C.}
\label{tab1}
\vspace{0.5pc}
\begin{tabular}{ccccc|c} 
$E_x$ (MeV) & $S_n$ (MeV) & $l_n$ & $J\pi$   & 
$S_{dp}\tablenote{From Ref. \protect\cite{Aj91}}$ 
&$\sigma_{n,\gamma}^{th}$ (mb) \\ \hline
   0.0  & 4.946 & 1   & $ 1/2- $ &  0.77    &    2.10           \\ 
  3.089 & 1.857 & 0   & $ 1/2+ $ &  0.65    &     -             \\ 
  3.684 & 1.262 & 1   & $ 3/2- $ &  0.14    &    1.10           \\ 
  3.854 & 1.093 & 2   & $ 5/2+ $ &  0.58    &     -             \\ \hline 
\multicolumn{5}{ c|}{Total}             &    3.20           \\ \hline
\multicolumn{5}{ c|}{Experimental}      &  3.53 $\pm$ 0.07 
\tablenote{From Ref. \protect\cite{Mux81}} \\
\end{tabular}
\end{center}
\end{table}


\begin{references}

\bibitem{Otx94} T. Otsuka, M. Ishihara, N. Fukunishi, T. Nakamura 
and M. Yokoyama, 
Phys. Rev. C {\bf 49}, R2289 (1994).


\bibitem{Nax91} Y. Nagai, M. Igashira, N. Mukai, T. Ohsaki, F. Uesawa, 
K. Takeda, T. Ando, H. Kitazawa, S. Kubono and T. Fukuda, 
Ap. J. {\bf 381}, 444 (1991).

\bibitem{Ohx94} T. Ohsaki, Y. Nagai, M. Igashira, T. Shima, K. Takeda, 
S. Seino and T. Irie, 
Ap. J. {\bf 422}, 912 (1994).

\bibitem{Igx95} M. Igashira, Y. Nagai, K. Masuda, T. Ohsaki 
and H. Kitazawa,
Ap. J. {\bf 441}, L89 (1995).

\bibitem{Nax94} T. Nakamura {\it et al.}, 
Phys. Lett. B {\bf 331}, 296 (1994).


\bibitem{Tax85Ta91} I. Tanihata, H. Hamagaki, O. Hashimoto, Y. Shida, 
N. Yoshikawa, K. Sugimoto, O. Yamakawa and N. Takahashi, 
Phys. Rev. Lett. {\bf 55}, 2676 (1985); 
I. Tanihata, 
Nucl. Phys. {\bf A522}, 275c (1991).


\bibitem{Kix94} J. D. King, R. E. Azuma, J. B. Vise,
J. G\"{o}rres, C. Rolfs, H. P. Trautvetter and A. E. Vlieks,
Nucl. Phys. {\bf A567}, 354 (1994); references therein.


\bibitem{Rax85} S. Raman, R. F. Carlton, J. C. Wells, E.T. Jurney,
and J. E. Lynn, 
Phys. Rev. C {\bf 32}, 18 (1985). 


\bibitem{Max83} G. J. Mathews, A. Mengoni, F. K. Thielemann
and W. A. Fowler, 
Ap. J. {\bf 270}, 740 (1983).

\bibitem{Wix90}
M. Wiesher, J. G\"{o}rres and F. K. Thielemann,
Ap. J. {\bf 363} 340 (1990).


\bibitem{Th51} 
R. G. Thomas, 
Phys. Rev. {\bf 84}, 1061 (1951). 
[This is the oldest reference that we have 
found on the subject.]

\bibitem{MI60} H. Morinaga and C. Ishii,
Prog. Theor. Phys. {\bf 23}, 161 (1960). 

\bibitem{LL60} A. M. Lane and J. E. Lynn, 
Nucl. Phys. {\bf 17}, 563 (1960); 
{\bf 17}, 686 (1960).


\bibitem{Ly68} J. E. Lynn, 
{\it The theory of neutron resonance reactions} 
(Clarendon Press, Oxford, 1968).

\bibitem{LM74} A. M. Lane and S. F. Mughabghab, 
Phys. Rev. C {\bf 10}, 417 (1974). 

\bibitem{Mex94} A. Mengoni, T. Otsuka and M. Ishihara,
in {\it Proceedings of a Specialists' Meeting on Measurement, 
Calculation and Evaluation of Phototn Production Data, 
Bologna, 1994}, edited by C. Coceva, A Mengoni and A. Ventura, 
(NEA/DOC/95/1), p. 185.

\bibitem{Lyx87} J. E. Lynn, S. Kahane, and S. Raman, 
Phys. Rev. C {\bf 35}, 26 (1987). 

\bibitem{Aj91} F. Ajzenberg-Selove,
Nucl. Phys. {\bf A523}, 1 (1991).

\bibitem{Mux81} S. F. Mughabghab, M. Divadeenam and N. E. Holden, 
{\it Neutron Cross Sections}, (Academic, New York, 1981), Vol. 1. 

\end{references}
\end{document}